\documentclass[aps,prb,twocolumn,superscriptaddress,oneside,floatfix,amsmath,showpacs,amssymb]{revtex4}
\usepackage{graphicx}
\usepackage{dcolumn}
\usepackage{bm}
\usepackage[usenames]{color}
\usepackage{doi}
\usepackage{hyperref}
\begin{document}

\bibliographystyle{unsrt}

\newcommand{\etal}{\textit{et al.}}

\title{Multi-orbital physics in lithium-molybdenum purple-bronze: going beyond paradigm.}

\author{Piotr Chudzi\'{n}ski}
\email{P.M.Chudzinski@uu.nl}
\affiliation{Institute for Theoretical Physics, Utrecht University}

\date{\today}


\begin{abstract}

We investigate the role of inter-orbital fluctuations in the low energy physics of a quasi-1D material - lithium molybdenum purple bronze (LMO). It is an exceptional material that may provide us a long sought realization of a Tomonaga-Luttinger liquid (TLL) physics, but its behaviour at temperatures of the order of $T^*\approx 30$K remains puzzling despite numerous efforts. Here we make a conjecture that the physics around $T^*$ is dominated by multi-orbital excitations. Their properties can be captured using an excitonic picture. Using this relatively simple model we compute fermionic Green's function in the presence of excitons. We find that the spectral function is broadened with a Gaussian and its temperature dependence acquires an extra $T^1$ factor. Both effects are in perfect agreement with experimental findings. We also compute the resistivity for temperatures above and below critical temperature $T_o$. We explain an upturn of the resistivity at 28K and interpret the suppression of this extra component of resistivity when a magnetic field is applied along the conducting axis. Furthermore, in the framework of our model, we qualitatively discuss and consistently explain other experimentally detected peculiarities of purple bronze: the breaking of Wiedmann-Franz law and the magnetochromatic behaviour. Our model consistently explains all these.

\end{abstract}

\pacs{}

\maketitle

\section{Introduction}

A search for materials with one dimensional character has been
always an exciting endeavour fuelled by a promise of observing
a highly correlated electron liquid with fully predicable properties, the Tomonaga-Luttinger liquid (TLL). On
the other hand it has been unrewarding because, either a dimensional
cross-over or an ordering driven by the Peierls transition, blocked our way to the exotic physics.
Purple bronze, Li$_{0.9}$Mo$_{6}$O$_{17}$ (LMO), has been always considered a great
promise\cite{McCarroll-first}: its exceptionally anisotropic band
structure and negligible effects of phonons indeed seemed to be a
perfect starting point. Indeed extremely anisotropic transport
coefficients\cite{Greenblatt-old-res-anis, Mandrus-optics,
Santos-transport-weird, Hussey-rhoB} down to temperatures of order
10K and remarkable signatures of TLL spectral
function\cite{JWAllen-old, Hager-bron-STM, JWAllen-1D+2D,
JWAllen-alphaRG} were clearly\cite{JWAllen-high-bulk,
JWAllen-growth-meth, JWAllen-comment-Greenblatt} detected in
experiments. However at the same time the promise has become a
challenge, because the physics around a presumed dimensional
cross-over around $T^*\approx 30$K turns out to be very unusual. There is a
mysterious resistivity upturn\cite{Greenblatt-old-res-anis,Hussey-rhoB} plausibly indicating a transition
into insulating state, but with very weak signatures of phase
transition\cite{Mandrus-optics} and absolutely no effect on the lattice (except
anomalous negative volumetric expansion constant\cite{Santos-thermal-expan}). All these
properties are susceptible to magnetic field\cite{Hussey-rhoB} while, at the same
time, magnetic susceptibility is well below the Pauli limit and
is dominated by paramagnetic fluctuations\cite{Mandrus-optics,Chakhalian-muons}. Moreover, most recent
measurements showed intriguing deviations from a standard 1D
physics already at higher temperatures $T\sim 150K$: anomalous ARPES scaling\cite{JWAllen-Tscal}
and broadening\cite{Dudy-ARPES} below energy scales of order 150K and breaking of
Wiedman-Franz law at the same energy scales\cite{Hussey-thermo} (again with no
influence of phonons).

On the theory side it has been established that the system can be considered as 1D two leg ladder\cite{Popovic-bron-DFT,Thomas-bronze} very close to the Mott transition\cite{Piotr-bronze}\cite{Nuss-DMFT}. Hence, from a very fundamental perspective, we face not only the problem of dimensional cross-over\cite{giamarchi_book_1d, Berthod-dim-cross, Raczkowski-dim-cross} but also a weakly doped Mott-insulator. This makes the problem even more challenging, but also much more exciting, since the many-body excitations (and spectral gaps) in the vicinity of the Mott state are currently hotly debated issues\cite{doped-Mott-rev, doped-Mott,Civelli-doped-Mott,Gull-doped-Mott-DMFT}. In the particular context of purple bronze our task is to construct a many body mechanism, consistently on the top of the well established 1D description, which could explain experimental phenomenology observed at energy scales from 200K down to the possible dimensional crossover at $T^*\approx 30$K. A crucial hint comes from a bit forgotten experimental work Ref.\onlinecite{Choi-colors}, where a magneto-chromatic effect was observed. The characteristic energy scale, a characteristic magnetic field when color changes, falls quite close to $T^*$. Authors' of  Ref.\onlinecite{Choi-colors} concluded that this must be manifestation of some electronic re-organization within the d-manifold, an effect inaccessible in a purely 1D model, where only one, the $d_{xy}$-orbital, was taken into account\cite{Piotr-bronze}. Extending the model by including an orbital degree of freedom makes a lot of sense also if one realizes\cite{Canadell-DFT-old} that in LMO any perpendicular hopping must be transferred through d-orbitals other than $d_{xy}$, so any out-of-1D processes, any dimensional cross-over, must involve excitations involving orbital-swap processes. Moreover, from the DFT studies\cite{Thomas-bronze} it is known that there is a lot of entropy available in structural fluctuations if we allow the system to explore the other d-orbitals.

The outline of this paper is as follows. In Sec.\ref{sec:model} we first introduce in detail a possibility for excitonic states and their interactions in LMO, then we write down the hamiltonian of the problem and explain physics covered by each of its constituents. To be precise, in our construction we start with a robust TLL, as found in Ref.\onlinecite{Piotr-bronze}, and then introduce a new "UV" cut-off at energy scale that is slightly larger than the spin-orbit coupling. The orbital-fluctuation effects enter into problem (as e.g. modified scattering amplitudes) at this energy which justifies the choice of the UV cut-off. Based on this input, in Sec.\ref{sec:observab} we compute observable quantities: the fermionic spectral function and electrical resistivity. This serves to asses the validity of our model. Then in Sec.\ref{sec:discuss} we discuss further experimental hallmarks of a presence of the excitonic physics.

\section{Model}\label{sec:model}

\subsection{Inter-orbital excitations}\label{ssec:inter-orbit}

\begin{figure}
\centering
  \includegraphics[width=\columnwidth]{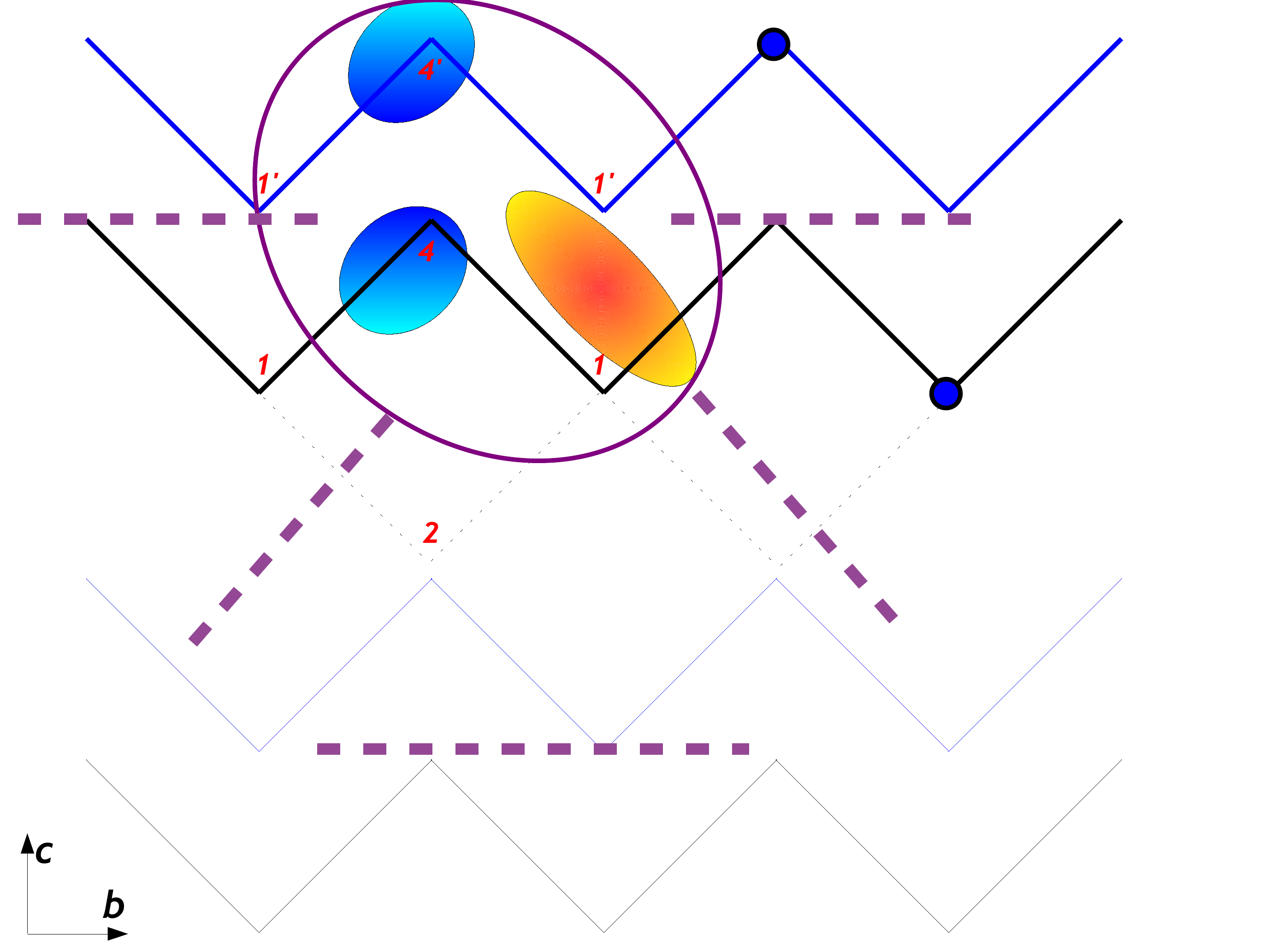}
\caption{Schematic illustration of plausible wavefunction of the
small exciton drawn on the top of b-c plane cross-section of the
LMO crystal structure. Only these Mo sites that host $t_{2g}$
electronic density are indicated. They are numbered according to notation in
Ref.\onlinecite{Canadell-DFT-old}. Pairs of zig-zag chains are
along the b-axis. Exciton with an average radius $r_{ex}$ is
indicated by a purple ellipse. Orange and blue ellipses show hole
in the bonding and electron in the anti-bonding bands
respectively. Dashed purple lines shows hopping paths in agreement
with Ref.\onlinecite{Nuss-DMFT}.}\label{fig:struct}
\end{figure}

Crystal structure of LMO is known\cite{Onoda-structure-Xray,
Santos-structure-neutrons} to consist out of 2D slabs parallel to
the b-c plane. Within the slabs pairs of quasi-1D zig-zag Mo
chains, which run along the b-axis, can be distinguished (see
Fig.\ref{fig:struct}). Each Mo atom is located inside an
octahedron built out of oxygen atoms, so due to a strong crystal
field split the low energy manifold is given by $t_{2g}$
orbitals\cite{Popovic-bron-DFT, Canadell-DFT-old, Thomas-bronze}.
The $e_g$ orbitals are $\sim$eV away from the Fermi energy $E_F$
and are projected out. Hence the $t_{2g}$ states can be treated
as a triplet with an effective $\tilde{L}=1$,  where
$d_{xz},d_{yz}$ are linear combinations of $\tilde{L}_z=\pm 1$. Since the split between the bare levels, favouring $\tilde{L}_z=0$, is tiny\cite{Thomas-bronze} ($\sim 10meV$) then the proper many-body description of LMO, the base for all considerations of this material, has to be given in terms of the following multi-orbital hamiltonian:
\begin{widetext}
\begin{multline}\label{eq:ham-tot}
H_{bs}= \sum_{\bar{\sigma},\alpha,k} \varepsilon_{\bar{\sigma},\alpha}(\vec{k})c_{\bar{\sigma}\alpha}^{\dag}(\vec{k})c_{\bar{\sigma}\alpha}(\vec{k})+U \sum_{\bar{\sigma}\alpha,r} c^{\dag}_{\bar{\sigma}\alpha}(\vec{r})c^{\dag}_{-\bar{\sigma}\alpha}(\vec{r})c_{-\bar{\sigma}\alpha}(\vec{r})c_{\bar{\sigma}\alpha}(\vec{r})+(U-2 J_H)\sum_{\bar{\sigma}\alpha,\beta,r}c^{\dag}_{\bar{\sigma}\alpha}(\vec{r})c^{\dag}_{\bar{\sigma}\beta}(\vec{r})c_{\bar{\sigma}\beta}(\vec{r})c_{\bar{\sigma}\alpha}(\vec{r})+\\
(U-J_H)\sum_{\bar{\sigma}\alpha,\beta,r}c^{\dag}_{-\bar{\sigma}\alpha}(\vec{r})c^{\dag}_{\bar{\sigma}\beta}(\vec{r})c_{\bar{\sigma}\beta}(\vec{r})c_{-\bar{\sigma}\alpha}(\vec{r})+J_H\sum_{\bar{\sigma}\alpha,\beta,r}c^{\dag}_{-\bar{\sigma}\alpha}(\vec{r})c^{\dag}_{\bar{\sigma}\beta}(\vec{r})c_{-\bar{\sigma}\beta}(\vec{r})c_{\bar{\sigma}\alpha}(\vec{r})+\sum_{\bar{\sigma}\alpha\beta\gamma,r,r'} V_{\gamma\delta}^{\alpha\beta}(\vec{r}-\vec{r'})c^{\dag}_{\alpha}(\vec{r})c^{\dag}_{\beta}(\vec{r'})c_{\gamma}(\vec{r'})c_{\delta}(\vec{r})
\end{multline}
\end{widetext}

where $c_{\bar{\sigma}\alpha}^{\dag}(\vec{k})$ is a creation operator of a fermion
with spin $\bar{\sigma}$, in the band $\alpha=d_{xy},d_{xz},d_{yz}$ with a momentum $\vec{k}$ and energy $\varepsilon_{\bar{\sigma},\alpha}(\vec{k})$, $c_{\bar{\sigma}\alpha}^{\dag}(\vec{r})$ is Fourier transform to real space. The last three terms are the strong correlations in the form of Hubbard term between electrons on the same orbital $(U+J_H)$, on different orbitals $U,(U-J_H)$, orbital exchange $J_H$ and the long range interactions term $V_{\gamma\delta}^{\alpha\beta}(\vec{r}-\vec{r'})$ (the 1D system itself is unable to fully screen Coulomb interactions\cite{giamarchi_book_1d}). The Hubbard terms are the largest energy scale in the problem\cite{Piotr-bronze,Popovic-bron-DFT,Nuss-DMFT} while $V_{\gamma\delta}^{\alpha\beta}(\vec{r}-\vec{r'})$ does depend on orbital index because of an extended nature of \emph{eigen-}wavefunctions. This constitutes a very complicated problem, whose solution is not accessible neither in analytic nor in numerical way, but even the $H_{bs}$ does not exhaust entire problem we face in the LMO. Since the LMO is likely to be very close to the quantum phase transition \cite{Piotr-bronze}, and we are interested in the low energy phenomena, we must incorporate further perturbations to obtain the full hamiltonian:      
\begin{equation}\label{eq:ham-tot2}
	H_{tot}=H_{bs}+\sum_{\alpha,k} t_{\alpha-\beta}(c_{\bar{\sigma}\alpha}^{\dag}(\vec{k})c_{\bar{\sigma}\beta}(\vec{k})+h.c.)+H_{so}
\end{equation}
These are inter-orbital hybridization and spin-orbit coupling term, where the latter is in a usual form $H_{so}\sim \Delta_{so}\vec{\hat{L}}\cdot\vec{\bar{\sigma}}$ (see Sec.\ref{ssec:coupl-def} for an estimate of $\Delta_{so}$). They are expected to be at least order of magnitude smaller than the smallest energy scale in $H_{bs}$ hence it is justified to consider them as perturbations and to zero order assume that spin and orbital are still good quantum numbers in our problem (so e.g. they can be used as an index of $\varepsilon_{\bar{\sigma},\alpha}(\vec{k})$). Obviously the extra terms are present only in a mutli-orbital version of our problem, actually they are unavoidable in LMO which is a system with a reduced symmetry (distorted octahedra) and for instance $J_H$ is not rotationally invariant in the orbital space (hence a tensor $\hat{J}_H$ shall generate the above given perturbations upon contractions of higher-order interaction terms). 

From previous DFT studies\cite{Popovic-bron-DFT, Canadell-DFT-old, Thomas-bronze} we know the single-particle dispersions $\varepsilon_{\alpha}(\vec{k})$ which tell us that within the $t_{2g}$ manifold, (a pair of) 1D-$d_{xy}$ bands crosses $E_F$. Actually, this assertion is confirmed by several ARPES experiments\cite{Dudy-ARPES}. Hence a foundation of the theoretical description of LMO has to be a many-body 1D model (so first one tackles Eq.\ref{eq:ham-tot} with $\alpha$ (orbital index) set to $d_{xy}$). Indeed, this is exactly what was accounted for within the TLL 1D theory in Ref.\onlinecite{Piotr-bronze}. 

On a single-particle level using the 1D physics is perfectly admissible because within the
$d_{xy}$ manifold the chains are not hybridized down to energies below 10meV (they are coupled
only via long-range Coulomb interactions). However the $d_{xz},d_{yz}$ are
split away from $E_F$ only due to a small inter-chain
$\pi$-hybridization gap $\Delta_h\approx 0.4eV$ formed between the
doublet of chains. Please note that this gap is not only smaller
than on-site Hubbard $U$, but also\cite{Blugel-cRPA-U} smaller
than Hund $J_H$, hence the states split by $\Delta_h$ are
susceptible to mixing by many body effects. This is particularly relevant for inter-orbital \emph{particle-hole} virtual excitations which must enter into a re-summation of Feynmann diagrams since they represent a sub-set of diagrams with the maximal divergence number. In other words, we aim to incorporate these $d_{xz},d_{yz}$ \emph{particle-hole} diagrams into the parquet re-summation that had so far lead us\cite{Piotr-bronze} to standard $d_{xy}$-TLL.  

In a dilute limit, the most natural way to re-introduce fermions from the $d_{iz}$
manifold back into the low energy sector is by postulating that
they can form an excitonic state with an energy close to the $E_F$.
Taking into account the size of the gap $\Delta_h$ (with $E_F$ in the middle of it), this needs to be a small
exciton with a binding energy of order 0.2eV. A standard argument,
opposing such construction, would be that even upon disregarding
1D metallic states from $d_{xy}$ we have a semiconductor with a rather
narrow gap, hence excitons should be large (in real space) with
small binding energies. However this overlooks a crucial
peculiarity of purple bronze, that comes from a proximity to the
Mott-CDW state: it is a propensity to form a chess-board pattern
of charges identified in Ref.\cite{Piotr-bronze} (see Fig.3b) therein) and later the tendency was confirmed in a
numerical DFT+DMFT study in Ref.\onlinecite{Nuss-DMFT}. This corresponds to an enhanced
charge susceptibility and indeed a direct calculation\cite{Merino-CDW-weird} within a
simpler RPA framework showed explicitly such enhancement with a
broad peak around $\vec{q}=(0,\pi/(b/2),\pi/(c/2))$. Staying on
the RPA ground, this implies an enhancement of an inverse
dielectric constant $\epsilon^{-1}(\omega=0,\vec{q}=(0,\pi/(b/2),\pi/(c/2)))$ which
enters to effective electron-electron interactions (we take $V_{\gamma\delta}^{\alpha\beta}$ in Eq.\ref{eq:ham-tot} which has screened Coulomb-character: $V_{\gamma\delta}^{\alpha\beta}=V_{Coul}\epsilon^{-1}$).
Then from result of Ref.\onlinecite{Merino-CDW-weird}, by taking a Fourier transform of $\epsilon^{-1}(q)$, we find that a pair of fermions located at a characteristic distance
$|r_{ex}|=\sqrt{(b/2)^2+(c/2)^2}$ can interact quite strongly.
This statement is valid no matter if the system actually reaches
the charge ordering\cite{Merino-CDW-weird} or only stays in the vicinity of it. We can then postulate an existence of a small exciton of a size $r_{ex}$,
where electron and hole are bound on a single plaquette consisting
out of four Mo sites. The plaquette is defined as a planar structure within a unit cell consisting of those Mo sites which contains, according to previous DFT studies\cite{Piotr-bronze,Popovic-bron-DFT,Nuss-DMFT}, a vast majority of the $t_{2g}$ density (see Fig.\ref{fig:struct}).  Since we are in a strong coupling limit, it
is very hard to estimate a binding energy of the exciton. Based on
a general argument, we only note that a hole (partially) located
in between the Mo chains (see Fig.\ref{fig:struct}) compensates an enhanced electronic density
formed by the $\pi-$bonding state\footnote{for an exciton of the
postulated size the bonding-band hole is constrained to be in that space}. So indeed, the bonding energy should be roughly of order of a half
of hybridization gap $\Delta_h$.

Due to an enlarged density of states at the band bottom exciton states appears in BZ (Brillouin zone) in between a maximum of valence and a minimum of conduction band. Based on the LMO band structure\cite{Thomas-bronze}, we know that in LMO for each $d_{iz}$ band, for $\pm k_c$, there are lines of valence band maxima and lines of conduction band minima that seem to be just above it. Hence for every $k_b$ there is a favourable condition (as a function of $k_c$) to create an intra-orbital direct exciton and an inter-orbital in-direct exciton. Summing up over $\vec{k}$ (see Eq.\ref{eq:a-def-2-quant}) allows to build a localized boson. Moreover, for such tight configuration of an electron and a hole, a relatively large Hund coupling on Mo atom may give an additional contribution to the binding energy. This is present for the inter-orbital exciton ($\tilde{L}=1$ configuration) and the electron-hole spin triplet configuration ($S=1$) and gives a favourable contribution from the spin-orbit coupling that can be as large as\citep{Iverson-Mo-LS} 0.1eV.
Resulting exciton is a so called “dark exciton” for which a direct recombination with emission of a photon is prohibited\footnote{triple photon process would be necessary, or re-combination due to phonons, but the recombination is blocked since phonons energies are small while $J_H$ is significant}.  We can then claim that a long-lived excitonic state with a sufficient binding energy is very plausible in LMO. We define an annihilation operator of such particle (for simplicity we put it at $x=0$):
\begin{multline}\label{eq:a-def-2-quant}
a(x=0)=\sum_{k_b}c_{d_{xz},\uparrow}(k_b)(z_{1}h_{d_{yz},\uparrow}(k_b)+z_{1}'h_{d_{yz},\downarrow}(k_b))+\\
c_{d_{yz},\downarrow}(k_b)(z_{2}h_{d_{xz},\downarrow}(k_b)+z_{2}'h_{d_{xz},\uparrow}(k_b))
\end{multline}
where $h_{\alpha,\bar{\sigma}}^\dag(k)$ is an operator that creates a hole in a band $\alpha$ with spin $\bar{\sigma}$ and momentum $k$, by definition $h_{\alpha,\bar{\sigma}}^\dag(k)\equiv c_{\alpha,-\bar{\sigma}}(-k)$. The $z_i$ are complex numbers, variational coefficients that need to be evaluated (in a low symmetry environment of LMO) by an independent calculation. The excitons live on b-c plane of LMO crystal and as such should be indexed by two coordinates. However, the core of this paper is dedicated to independent excitons appearing (or passing through) one chosen 1D TLL system. Some possible issues related to 2D excitons' interactions and the perpendicular dynamics shall be handled in Sec.\ref{ssec:excit-propert}.    

One can raise an issue: if relaxation into these states is so weak then how they get populated at all. Naturally, in any electronic liquid there is an incoherent background of excitations, they are of thermal, phononic or kinematic origin and constitute e.g. the screening cloud in the dilute limit. But these are mostly short lived, "bright" excitons. However if a $d_{xy}$ fermion interacts with an exciton, and as we show below it indeed does, then there is a possibility of a spin-exchange process ($\sim J_H$) with one of the exciton's constituents. Upon such exchange standard $S=0$ exciton turns into the $S=1$ dark exciton. Since the 1D $d_{xy}$ fermion can also absorb any perpendicular momentum $k_c$, then the indirect, inter-band excitons are created as well.
This scattering channel is particularly strong when the $k_b$ coordinate of maximum/minimum of  valence/conductance bands fall close to the $k_{F}$ of the 1D $d_{xy}$ fermions, since then the envelopes of respective Bloch waves are the same.  From LMO's band structure\cite{Popovic-bron-DFT,Thomas-bronze} we know that this condition is met along the P-K line of the 1st BZ [Actually, on this peculiar line velocities of $d_{iz}$ fermions (slope of band structure slightly away from extrema) are close to $V_F$ of $d_{xy}$. This further increases the scattering cross-section]. We then deduce that if our conjecture about the exciton's creation in exchange process is true then the "dark" excitons will be the most efficiently produced (and visible) along the P-K line.

We conclude this section by stating that the
way to go beyond the pure 1D model\cite{Piotr-bronze}, to capture the physics of Eq.\ref{eq:ham-tot2}, is to include excitonic
effects by introducing the following hamiltonian:
\begin{equation}
\tilde{H}_{tot}=H_{1D}+H_{ex}+H_{f-ex}
\end{equation}
where the term $H_{1D}$ describes 1D-$d_{xy}$ fermions (described in Sec.\ref{ssec:TLL-def}), the term $H_{ex}$ describes $d_{xz}$- $d_{yz}$ excitons postulated in Sec.\ref{ssec:inter-orbit} (we come back to it in Sec.\ref{ssec:excit-propert}) and the term $H_{f-ex}$ describes coupling between the 1D fermions and the excitons (studied in Sec.\ref{ssec:coupl-def}). The approximation $\tilde{H}_{tot}$ is valid for a specific case when only 1D bands are crossing Fermi energy while other $t_{2g}$ orbitals contribute through a stable, low energy exciton particles (one can call it the "exciton pole" approximation).

\subsection{Physics of $d_{xy}$ fermions}\label{ssec:TLL-def}

The 1D sector
$H_{1D}$ can be described by the Tomonaga-Luttinger
liquid\cite{Haldane-1Dmain, giamarchi_book_1d}, whose peculiar properties are well known. For the particular case of LMO the procedure of deriving TLL was outlined in detail in
Ref.\onlinecite{Piotr-bronze}. The TLL hamiltonian reads:

\begin{equation}\label{eq:ham-TLL-def}
    H^{1D}[\nu]= \sum_{\nu} \int \frac{dx}{2\pi}
    \left[(v_{\nu}K_{\nu})(\pi \Pi_{\nu})^{2}+\left(\frac{v_{\nu}}{K_{\nu}}\right)(\partial_{x} \phi_{\nu})^{2}\right]
\end{equation}

where $v_{\nu},K_{\nu}$ are velocity and TLL
parameter of a given bosonic mode $\nu$, these
depend on electron-electron interactions (terms $\sim U,V$ with $\alpha=d_{xy}$ in Eq.\ref{eq:ham-tot}) with small momentum
exchange. Formally Eq.\ref{eq:ham-TLL-def} is a result of a parquet re-summation of all $\alpha=d_{xy}$ terms in Eq.\ref{eq:ham-tot}. This part are the density-density interactions that can be expressed using $\rho_{\nu}(x)$ where e.g. $\rho_{\bar{\sigma}}(x)=c^{\dag}_{\bar{\sigma}}(x)\hat{\bar{\sigma}}_{\bar{\sigma},\bar{\sigma'}}c_{\bar{\sigma'}}(x)$ ($\hat{\bar{\sigma}}_{\bar{\sigma},\bar{\sigma'}}$ is a Pauli matrix acting in a spin space). The
Eq.\ref{eq:ham-TLL-def} is written in terms of density fields $\phi_{\nu}(x)$ and
canonically conjugate fields $\theta_{\nu}(x)$, with $\Pi_{\nu}(x)=\partial_{x}\theta_{\nu}(x)$. To define them,
first one extracts long wavelength behavior around the Fermi
points. $c_{\bar{\sigma}\alpha}^{\dag}(x)=\exp(i k_F
x)\psi_{R\bar{\sigma}\alpha}^{\dag}(x)+\exp(-i k_F
x)\psi_{L\bar{\sigma}\alpha}^{\dag}(x)$ and then introduces
bosonic fields, collective modes,
$\psi_{R/L,\bar{\sigma},\alpha}(x)=\kappa_{R/L\bar{\sigma}\alpha}\frac{1}{2\pi\alpha}\exp(i[\sum_{\nu}\bar{\sigma}\alpha(\phi_{\nu}(x)\pm\theta_{\nu}(x))])$
(where $\kappa_{R,L\bar{\sigma}\alpha}$ is a constant operator,
Majorana fermion, introduced to ensure proper anti-commutation
relations). A real space field $\psi_{\bar{\sigma}\alpha}(x)$ is an eigenvalue of the of the second quantization
operators $c_{\bar{\sigma}\alpha}^{\dag}(x)$ (that creates 1D $d_{xy}$ fermion in a given point), in the Fock space of the coherent states. In the case with two 1D $d_{xy}$ bands and thus four Fermi
points, a two-leg ladder model is necessary to
describe the physics\cite{giamarchi_book_1d}. Both spinon and holon acquire
total and transverse flavor, in
Eq.\ref{eq:ham-TLL-def}, $\nu = \rho\pm, \sigma\pm$.

As usual in 1D theory, TLL state is perturbed by
non-linear cosine terms. In Ref.\onlinecite{Piotr-bronze} we have found that the most relevant
cosine terms are the umklapp terms. These are most likely to drive us away from the
critical TLL state and open spectral gaps. They involve charge
field $\phi_{\rho+}$ and their relevance is intimately related to the long-range
character of interactions which drive $K_{\rho\pm}$ way below
the non-interacting value $K=1$. The non-linear terms are:

1. Umklapp scattering (at quarter filling): 

\begin{multline}\label{eq:umklapp-comb}
H_{um}=
g_3 \int dx \cos(4\phi_{\rho +}(x)+\delta x)\cdot [\cos(4\phi_{\rho -}(x))+\\
\xi\left\lbrace\cos(4\phi_{\sigma+}(x))-\cos(4\theta_{\sigma-}(x))+\cos(4\phi_{\sigma-}(x))\right\rbrace]
\end{multline}

where $\delta$ is a finite doping $\sim 1\%$ present in LMO. On the top of an intra-chain umklapp we have introduced, following Ref.\onlinecite{Gogolin-zigzag} , various inter-chain processes with amplitudes multiplied by a coefficient $\xi\ll 1$ and signs determined by the convention of Klein factors.

2. Inter-chain $4k_F$ processes:
\begin{equation}\label{eq:4kF}
    H_{4k_F}^{\pi} = g^{\pi}_{4k_{F}} \int dx [\cos(4\phi_{\rho+}(x)+\delta x)+\cos(4\phi_{\rho-}(x))]
\end{equation}

3. Further exchange scattering processes ($q=2k_F$) that are present in TLL irrespective of commensurabillity. In the case of two chains coupled by interactions the most relevant terms are:
\begin{equation}\label{eq:2kF}
    H_{1}^{\pi} = g^{\pi}_{2k_{F}} \int dx \cos(2\phi_{\rho-}(x))[\cos(2\phi_{\sigma +}(x))+\cos(2\theta_{\sigma -}(x))]
\end{equation}

4. Moreover there are the single particle cosine terms that push us outside TLL fixed point:
\begin{itemize}
\item the single particle hopping between the chains, which according to Ref.\cite{Khveshchenko-RGtperp} has the following form: $t_{\perp}\int dx [\cos(\phi_{\rho
-})\cos(\phi_{\sigma-})+\cos(\phi_{\rho+}+\delta x)\cos(\phi_{\sigma+})]\cos(\theta_{\sigma-})\cos(\theta_{\rho-})$ 

\item the potential backscattering on the Peierls distortion which may have several variants. In general it is a term: $V_{P}\int dx \exp(\phi_{\rho+}) \cos(\phi_{\rho
-})F[\cos(\phi_{\sigma+}),\cos(\phi_{\sigma-}),\cos(\theta_{\sigma-})\sin(\phi_{\sigma+}),\sin(\phi_{\sigma-}),\sin(\theta_{\sigma-})]$, where a functional $F[]$ is a linear combination whose precise form encodes the spin/chain dependence of backscattering\cite{PC-zigzagCNTs}.
\end{itemize} 
[Please note that in all formulas above we deal with the case of an interaction coupled ladder $V_{\perp}\gg t_{\perp}$ which is less common than the hybridization coupled ladder. In particular this shall cause a swap $\theta_{\rho-}\leftrightarrow\phi_{\rho-}$ in some cosines' arguments.]

In Ref.\onlinecite{Piotr-bronze} we studied, using standard RG methods, the TLL instabilities due to above given non-linear terms. It was found that in the range of parameters characterizing LMO, possible spectral gaps (if any) are smaller than 1meV. Hence the effects introduced in the following Sec.\ref{ssec:coupl-def} are not suppressed by any spectral gap, on the contrary they are stronger and should be introduced to theory as  perturbations on the level of Eq.\ref{eq:ham-TLL-def}.

\subsection{Fermion-exciton coupling}\label{ssec:coupl-def}

\subsubsection{The coupling}

We now introduce a coupling between 1D fermions and the excitons. In a small coupling limit we restrict ourselves to the lowest order possible coupling term which one can write down:
\begin{equation}\label{eq:f-ex-def}
H_{f-ex}=\sum_{\bar{\sigma}\bar{\sigma'},k,q} V_{f-ex}(k,q)(c_{\bar{\sigma}}(k)\hat{O}_{f-ex}c^{\dag}_{\bar{\sigma'}}(k-q)a^{\dag}_q+h.c.)
\end{equation}
where the exact form in the charge/spin/orbital space, the $\hat{O}_{f-ex}$, is to be derived in this section.  

Exciton is a neutral and tightly bound object, so its Coulomb-interaction with
an electron should be weak. However, there is a coupling with a spin
degree of freedom. It originates from two distinct sources:
\begin{itemize}
\item $H_{f-ex}^{(1)}$ is a coupling between an angular
momentum of an exciton $J$ and a spin $\bar{\sigma}$ of a $d_{xy}$ fermion. This is due to spin-orbit coupling and so:
\begin{equation}\label{eq:ham-ex-fer1}
H_{f-ex}^{(1)} \sim \hat{J}\cdot \hat{\bar{\sigma}}(x)
\end{equation}.

\item $H_{f-ex}^{(2)}$: a tightly bound electron-hole pair does
produce a large local electric field which can couple with a
moving electron via the Rashba effect:
\begin{equation}\label{eq:ham-ex-fer2}
H_{f-ex}^{(2)}\sim \hat{\bar{\sigma}}\times \hat{k}
\end{equation}
\end{itemize}

Aiming for a second quantization formula, Eq.\ref{eq:f-ex-def}, we define the coupling potentials e.g. $V_{f-ex}^{(1)}(k,q)=\langle\Psi'|\hat{J}\cdot \hat{\bar{\sigma}}(x)|\Psi\rangle$ , where $|\Psi\rangle$ is a state of an entire system (close to its ground state), a tensor product of TLL and excitons. The strength of such coupling, we call it $\gamma$ (so $\gamma\equiv max(|V_{f-ex}(k,q)|)$), can be
determined from the Hund's rules applied to 4-d $Mo-t_{2g}$
orbitals in the local, atomic $Mo-d_{ij}$ limit. When $d_{iz}$
orbitals are both occupied in low spin-state, then there is one unpaired $d_{xy}$ occupied (1/2 probability). This configuration is a local doublet $S^2$ (other states with $L=1$ are eV's away).
From this perspective a fluctuation to $d_{iz}$ manifold of
a conduction electron is favorable since the resulting $P^3$
state (favoured by Hunds rules) has energy lowered by -0.12eV, so $\Delta_{so}^{eff}\approx 0.12eV$. There are four Mo sites within one unit cell and wavefunctions can further delocalize (with probability $\approx 10\%$) onto outer Mo sites, so from a basic combinatorics a probability of such contact interaction is $\binom{4}{2} (\frac{1}{4})^2 0.9^2 \approx 0.1$. Hence a strength of interaction between exciton and $d_{xy}$-fermion is roughly 10\% of $\Delta_{so}$, that is $\gamma\approx 12meV$.

The second term is not usually encountered and we shall tackle it
in detail below. Here we only notice that while
angular momentum is related to kinetic energy of electron-hole
system, the electric field is related to a potential energy of the
same system. The two must be equal. Since, both Rashba and $\hat{J}\cdot
\hat{\bar{\sigma}}$ are of a relativistic origin then we expect both terms $H_{f-ex}^{(1,2)}$ to be approximately equal. 

The delocalized nature of LMO eigenstates, which is known from all DFT studies\cite{Popovic-bron-DFT, Canadell-DFT-old, Nuss-DMFT}, and a finite size of the excitons both imply that the fermion-exciton scattering process takes place along a non-negligible range of distances. Hence the $V_{f-ex}(x,x')$, where $x'$ is an electron-exciton distance, is certainly not a Dirac delta function $\delta(x')$. Then, upon Fourier transformation to momentum space, one realizes that $V_{f-ex}(k,q)$ is a decaying function of the exchanged momentum $q$, which justifies our focus on forward (density-density) scattering processes. For concreteness, if we take both wavefunctions (of an exciton and a fermion) to be Gaussian packets then their convolution is also a Gaussian with a width of order $\sqrt{2}b$. By taking a Fourier transform one obtains a Gaussian in $q$-space, hence a function slowly decaying at low q, but quickly suppressed for $q>\pi/(2b)$.   

The fermion-exciton coupling will renormalize (via ladder series of vertex corrections) the effective excitonic propagator\footnote{strictly speaking the exchange coupling of $\tilde{L_z}=0$ with $\tilde{L_z}=\pm 1$ states will be renormalized as well, but we do not focus on it in the following}. The bare energy $\varepsilon_0$ of an exciton may depend on momentum due to a bare hopping integral $\Gamma_{ij}$ (see Eq.\ref{eq:ham-excit} below) but we neglect it for a moment. In a mean field picture we integrate out the fast fermionic states and in a very similar way like in Ref.\onlinecite{Martin-acousticTLL} we arrive at:
\begin{equation}\label{eq:eff-prop-ex}
D_{ex}^{-1}(k,\omega)=\omega^2+\varepsilon_0^2+\frac{\gamma^4 k^4}{\omega^2+(v_{\sigma+}k)^2}
\end{equation}
where momentum is counted in units $\alpha$. In the limit of small $\omega$ and large $v_{\sigma+}$ Eq.\ref{eq:eff-prop-ex} leads to an effective propagator with a linear dispersion $\epsilon_{eff}(k) \sim (\gamma/\alpha) \gamma/(\alpha v_{\sigma+})\cdot k$ where $\alpha$ is a UV cutt-off in reciprocal space (taken the same for spin-orbit coupling and kinetic energy). Please note that according to Ref.\onlinecite{Piotr-bronze}, the $v_{\sigma+}\sim J_{eff}\approx 0.2eV$ (thus $v_{\sigma+}\ll V_F$), hence a perturbation $\gamma$ (as well as the induced velocity) is substantial at energies of order $T^*$.

\subsubsection{Bosonization}

Standard, textbook\cite{mahan_book}
way of dealing with a coupled fermion-boson problem Eq.\ref{eq:f-ex-def}, in a single-fermion limit, is by applying the Lang-Frisov transformation: $\hat{U}(k)=\exp(\hat{S}(k))$, with $\hat{S}^{(1,2)}(k)=\sum_{q'}n(x)\exp(\imath
q'x)\frac{V_{f-ex}^{(1,2)}(k,q')}{\varepsilon_0(q')}(a^{\dag}_{q'}+a_{-q'})$,
which absorbs the $H_{f-ex}^{(1,2)}$ term. This transformation corresponds to
dressing up a fermion with a “polaronic” cloud:
\begin{multline}\label{eq:Lang-Frisov}
\tilde{\psi_k}(x)=\\
\exp \left(
\sum_{q'}n(x)\exp(\imath
q'x)\frac{V_{f-ex}^{(1,2)}(k,q')}{\omega_{q'}}(a^{\dag}_{q'}
+a_{-q'}) \right)\psi_k(x)
\end{multline}
where $q'$ is a momentum absorbed/emitted by an exciton and $\omega_{q'}$ is its energy and $n(x)$ is a number of fermions at a given point, in the following we take $n\approx 1$ in a position of exciton creation. The knowledge of $\tilde{\psi_k}(x)$ in principle opens a way to compute the spectral function, although further approximation about the nature of electron-boson coupling is necessary (e.g. adiabatic regime approximation). The Lang-Frisov transformation contains rich phenomenology of polaronic
physics, but it is sufficient only for weakly interacting fermions.
On the contrary from several experiments\citep{Dudy-ARPES, Hager-bron-STM, JWAllen-Tscal, JWAllen-alphaRG} it is known that in purple bronze we are dealing with TLL dominated by interactions (the single particle exponent
$\alpha$ is significant, $\alpha>0.5$), which constitutes a much more
complicated problem where electron-electron interactions cannot be taken perturbatively.

To tackle this problem, we note that in our case the $V_{f-ex}(k,q')$ does not depend on $q'$ for small enough $q'$
and the bare exciton energy $\varepsilon_0(q')$ is approximately constant\footnote{more precisely, for a non-zero $\Gamma_{ij}$, one can take $q^2$ dependence both in numerator and in denominator, that cancel out each other}. Then a
remaining summation in Eq.\ref{eq:Lang-Frisov}, of operators $a_{q'}$ over $q'$, gives us a purely local
operator $a(x)$. Since $V_{f-ex}^{(1,2)}$ have forward character (Sec.\ref{ssec:coupl-def}), and a large exchanged momentum could destroy the exciton, we restrict only to
processes when fermion and boson exchange small momenta $q'$. So from now on we take interaction term of only one argument $V_{f-ex}^{(1,2)}(k,q=0)$.  With
this assumption we can focus on the low momentum sector of 1D theory that can be expressed in terms of fermionic ($d_{xy}$) density operators $\rho_{\nu}(x)$. Then, not only the local
polaronic cloud factorizes out, but furthermore, thanks to
linearity $V_{f-ex}(k)\sim k$ each contribution to two particle
density $\tilde{\rho}_q=\sum_k \tilde{\psi}_k^{\dag}\tilde{\psi}_{k-q}$ shall have
the same factor: $\tilde{\rho}_q=\exp(q V_{f-ex})\rho_q$. Thanks to that we can separate out
“polaronic” contribution and proceed with a standard
construction of TLL based on the $\rho_q$ bosonic operators
multiplied by $\exp(\hat{S}(q))\sim \exp(cste\cdot q)$. In particular this means that entire Sec.\ref{ssec:TLL-def} remains valid also in the presence of bosons and we can go ahead with bosonization of $H_{f-ex}^{(1,2)}$

Once it is established that TLL construction holds, and that (also in the presence of interactions) the coupling has a \emph{local} character of excitons with a density of 1D $d_{xy}$ fermions, then the $H_{f-ex}^{(1)}$ term can be bosonized in a rather straightforward way as
it is equivalent to a displacement coupling between an amplitude of a
spin-wave $\hat{\bar{\sigma}}(x)$ and the \emph{local} angular momentum $J$. Since, by construction of the excitonic state, $\langle J_z \rangle \approx 0$, then we may focus on the transverse component.
Then the coupling reads:
\begin{equation}\label{eq:ham-disord-1}
H_{f-ex}^{(1)}= \int dx \gamma \nabla\phi_{\sigma+}(x)|J|(a^{\dag}(x)+a(x))
\end{equation}
where we used the fact that spin density $\rho_{\bar{\sigma}}(x)$ is related to a gradient of a spinfull collective mode $\rho_{\bar{\sigma}}(x)=\nabla\phi_{\sigma+}(x)$, and the translational symmetry of this term (so $V_{f-ex}(x)=\gamma=cste$). 

For $H_{f-ex}^{(2)}$ the coupling is with momentum density. Based on a form of the Rashba coupling $\hat{H}\sim \hat{\bar{\sigma}}\times
\hat{k}$ and the fact that electric field changes sign when we
swap the legs of the ladder, we can make a conjecture that the
bosonized version of the 1D hamiltonian will be:
\begin{equation}\label{eq:ham-disord-2}
    H_{f-ex}^{(2)}= \int dx \gamma (a^{\dag}(x)+a(x)) \Pi_{\sigma-}(x)
\end{equation}
where $\Pi_{\sigma-}$ is a momentum operator of a spin $\sigma-$
mode, that is a relative spin fluctuation between the two chains.
Since $\Pi_{\sigma-}(x)=\nabla\theta_{\sigma-}(x)$,
Eq.\ref{eq:ham-disord-2} (as well as Eq.\ref{eq:ham-disord-1}), is very similar to a problem with a forward
scattering on local “impurities”. This is not an accidental coincidence: excitons are in an extremely dilute limit and, above $T_o$, their relative positions and angular
momenta are uncorrelated, so a good approximation is that
the $d_{xy}$ fermions randomly creates bosons, drags them and then
release.  However, there is also a difference: in comparison with the
standard case of a varying chemical potential, the $\phi_{\rho +}$
mode is substituted with $\theta_{\sigma-}$ mode, that expresses a
kinematic character of the $H_{f-ex}^{(2)}$ coupling.

Overall, the hamiltonian describing gapless TLL plus exciton reads:
\begin{multline}\label{eq:ham-disord-f}
   \tilde{H}_{tot}=H_{TLL} + \int dx \gamma (a^{\dag}(x)+a(x)) \nabla\theta_{\sigma-}(x)+\\
    \int dx \gamma (a^{\dag}(x)+a(x)) \nabla\phi_{\sigma+}(x)+H_{\tilde{ex}}
\end{multline}

Based on considerations in Sec.\ref{ssec:excit-propert}, above the $T_o$ an effective hamiltonian for excitons (given momentarily in its 2D form) is $H_{\tilde{ex}}=\sum_{\vec{k}} [c_{\tilde{ex}} k_b a_{k_b,k_\perp}^{\dag}a_{k_b,k_\perp} + \Gamma k_{\perp}^2  a_{k_b,k_\perp}^{\dag}a_{k_b,k_\perp}]$ with an effective velocity along the b-axis determined by the self-energy correction $c_{\tilde{ex}}=\gamma (\gamma/v_{\sigma+})$. However, since the hopping $\Gamma$ is the smallest energy scale in the problem, order of magnitude smaller than $V_{f-ex}^{(1,2)}$, taking the adiabatic approximation for excitons is justified. In Eq.\ref{eq:ham-disord-2} we observe that the two terms $H_{f-ex}^{(1)}$ and $H_{f-ex}^{(2)}$
involve different bosonic modes, so in bosonic language it becomes
clear that they commute and e.g. one can define the Lang-Frisov
transformations for each of them separately.


\subsection{Properties of excitons}\label{ssec:excit-propert}

The excitonic part of the system $H_{ex}$ can be described as a
system of bosons (tightly bound small excitons) moving on a
triangular lattice (Fig.\ref{fig:struct}) with a bare hopping integral $\Gamma_{ij}$. $\Gamma_{ij}$ is a correlated hopping of an electron and a hole for which one must pay an energy cost $\Delta_h$. It is driven by an inter-orbital hybridization $t_{xz-xy}$ on the Fermi level. $t_{xz-xy}$ must be of the same order as a wiggling of Fermi surface found in DFT calculations, that is $\sim 15meV$. From simple perturbation theory one finds: $\Gamma_{ij}=4t_{xz-xy}^2/\Delta_h \approx 2meV$ where a factor four accounts for different ways of performing the process in spin-orbital space.

Up to now we neglected 2D interactions between excitons. This is a significant deficiency, since these are hard-core bosons with strong
interactions, that is putting two excitons on the same plaquete has an
energy cost $U_{ex}$ of a few eV ($\approx U$ in Eq.\ref{eq:ham-tot}) and on the neighbouring plaquetes
$V_{ex}(r=4b)$ is of order 0.5eV ($\approx V(2b)$ in Eq.\ref{eq:ham-tot}), the latter value is found from estimates in Ref.\onlinecite{Piotr-bronze,Popovic-bron-DFT}. Overall the $H_{ex}$ reads:
\begin{multline}\label{eq:ham-excit}
H_{ex}= \sum_{i,j} \Gamma_{ij}(a^{\dag}(i)a(j)+h.c.) + \sum_{i} U_{ex}/2 n_a(i) (n_a(i)-1) +\\
\sum_{i,r} V_{ex}(r)n_a(i)n_a(i+r)+ \mu \sum_{i} n_a(i) 
\end{multline}
where $n_a(i)$ is a boson density on an \emph{i-}th plaquette, $\mu$ is their chemical potential. 
Interactions between further plaquettes
$V_{ex}(r)$ are included.
On a-c plane these are weakly screened dipole-dipole interactions
decaying as $1/r^3$ while along b-axis these are retarded interactions mediated by spinons, second order in $\gamma$. However, since the bare excitons' hopping is tiny, even when $r=20b$ (large dilutions) interactions can be dominant that is $V(20b)>\Gamma_{ij}$. From DFT\cite{Popovic-bron-DFT} we know that an optimal admixture of $d_{xz},d_{yz}$ is of order $\sim 1\%$, so we expect exciton-exciton distances to be of order $\delta_{ex}^{-1}\approx 10b$ (where $\delta_{ex}$ is a dilution of an excitonic liquid).

Overall, we can distinguish three distinct regimes of excitonic physics:
\begin{description}
\item[$(T,\omega)>\gamma$] at highest energies "dark excitons" are very rare and decoupled from the rest of the system
\item[$(T,\omega)<\gamma$] in this \emph{intermediate} regime "dark excitons" can be efficiently created and scattered by $d_{xy}$ fermions, these are random processes and excitons can be considered bound to $d_{xy}$ fermions
\item[$(T,\omega)< \gamma^2 V_{ex}(r=\delta_{ex}^{-1})$] in this lowest energy regime the long range excitons' interactions play a dominant role
\end{description}

It should be noted that, when the system passes from the \emph{intermediate} to the low temperature regime, the renormalized dispersion of the hard-core bosons $\epsilon_{eff}(k)$, see Eq.\ref{eq:eff-prop-ex}, may cross $E_F$ and then bosonic system acquires a finite chemical potential.

Dilute gas of hard-core bosons can be mapped, using
Holstein-Primakoff transformation, onto a (pseudo-)spin system:
$\tilde{\tilde{S}}_+(x)=a^{\dag}(x)$,$\tilde{\tilde{S}}_-(x)=a(x)$,$\tilde{\tilde{S}}_z(x)=a^{\dag}(x)a(x)-1/2$. $\tilde{\tilde{S}}_z$ is related to an occupation (or not) of a
given plaquette by an exciton. The hamiltonian Eq.\ref{eq:ham-excit} can be rewritten in terms of $\tilde{\tilde{S}}$ operators:
\begin{equation}\label{eq:ham-tilde-S}
H_{ex}= \sum_{i,j} \Gamma_{ij}\tilde{\tilde{S}}_+(i)\tilde{\tilde{S}}_-(j) + \sum_{i,r} J_z(r)\tilde{\tilde{S}}_z(i)\tilde{\tilde{S}}_z(i+r)
\end{equation}
where $J_z(r)\equiv V_{ex}(r)$ is a long range uni-axial (pseudo-)spin-coupling. By taking into account
the density-density character of repulsive interactions in
Eq.\ref{eq:ham-excit}, we can focus on the last term in Eq.\ref{eq:ham-tilde-S} and the problem is simplified to the 2D
anti-ferromagnetic Ising model on a triangular lattice (see Fig.\ref{fig:struct} where hopping paths of excitons, that define connected sites of the pseudo-spin lattice, are indicated). Since
$U_{ex}$ and $V_{ex}(r=2b)$ are by far the largest energy scales,
such model can be solved by building larger and larger blocks, with one $\tilde{\tilde{S}}_z=+1/2$ inside, that
are anti-ferromagnetically ($\tilde{AF}$) coupled with their vicinity (in the original language this corresponds to Wigner crystal formation). Clearly we can continue this procedure as long as $V_{ex}(r_b)\geq\Gamma_{ij}$, where $r_b$ is the block size. This is a
strongly frustrated model and is known\cite{Moessner-QIsing} to undergo an
order-by-disorder phase transition favouring the most flippable $\tilde{AF}$ state. The transition is driven by an entropy gained due to
quantum tunneling between up/down pseudo-spin states $\Gamma_{ij}$ (that is a
hopping of excitons in our original model). A critical temperature
$T_o$ is a fraction of the hopping\cite{Moessner-QIsing}, roughly $\Gamma_{ij}/2$. From our previous study on LMO we know that this is of the same order as experimentally observed crossovers at 25K,
which suggest to interpret the experimentally observed $T^*$ as
such transition. This conjecture is supported by a very weak
signature of the transition in the specific heat and lack of a signal in X-ray scattering (as expected for a BKT phase\cite{Moessner-QIsing} of neutral bosons). 



\section{Observables}\label{sec:observab}

To validate our construction we calculate some of its
experimentally observable consequences. We begin, in
Sec.\ref{ssec:spectr-fun}, in the intermediate energy regime above $T_o$ where
the physics is defined by 1D fermions (TLL) randomly scattering on
excitons (with scattering amplitude $\sim$10meV). The non-linear
terms, Eq.\ref{eq:umklapp-comb}-Eq.\ref{eq:2kF}, are ineffective in this regime so we disregard any spectral
gaps($\sim$1meV) and take Eq.\ref{eq:ham-TLL-def} as a good
approximation for the $d_{xy}$ fermions.

\subsection{Spectral function}\label{ssec:spectr-fun}

The first observable, which we want to compute, is the spectral function of 1D fermions which is directly measurable by probes such as ARPES or STM. The presence of
excitons modifies fermionic spectral function. We follow a case of
a forward scattering on disorder\cite{TG+HS-disord} and absorb fermion-exciton coupling terms in Eq.\ref{eq:ham-disord-f} by phase shifts of bosonic $\theta_{\sigma-}$ and $\phi_{\sigma+}$ modes, e.g.:
$\tilde{\theta}_{\sigma-}=\theta_{\sigma-}+\frac{K_{\sigma-}^{-1}}{v_{\sigma-}}\int^x
dx' \gamma (a^{\dag}(x')+a(x'))$. An important modification in
comparison with the studies of a forward disorder is that instead of random
immobile impurities (that do not conserve momentum) now we study
random scattering on a coherent bosonic bath, an effect that does enter to correlation function. From bosonization basics, outlined in Sec.\ref{ssec:TLL-def} we know that the fermionic fields $\psi(x)$  depend on bosonic fields as $\exp\imath(\sum_\nu\phi_{\nu}+\tilde{\theta}_{\nu})$. Clearly, the functional dependence on bosonic operators is the same as in Eq.\ref{eq:Lang-Frisov} which not only justifies our approach but also allows to note that the hermitian dynamics of polarons will translate into a proper commutation relations of $\tilde{\theta}(x)$ fields (same commutation algebra as for the standard TLL $\theta(x)$ field). On the level of fermionic correlation function the shift $\theta_{\sigma-}\rightarrow\tilde{\theta}_{\sigma-}$ can be accommodated by:
\begin{widetext}
\begin{multline}
\langle\exp\imath(\tilde{\theta}_{\sigma-}(x,t)-\tilde{\theta}_{\sigma-}(0,0))\rangle=\\
\left\langle\exp\imath\left[\frac{K_{\sigma-}^{-1}}{v_{\sigma-}}\int^x
dx' \gamma (a^{\dag}(x')+a(x'))\right]\exp-\imath\left[\frac{K_{\sigma-}^{-1}}{v_{\sigma-}}\int_x^0
dx'' \gamma (a(x'')+a^{\dag}(x''))\right]\right\rangle \langle\exp\imath(\theta_{\sigma-}(x,t)-\theta_{\sigma-}(0,0))\rangle
\end{multline}
\end{widetext}
where we assume the first order coupling between exciton and TLL, such that correlation function factorize into TLL and excitonic parts. Furthermore, taking the \emph{adiabatic} limit for bosons implies that there is no \emph{time-dependence} imposed by the presence of the a-operators. Since only the connected diagrams (also for excitons-fermions processes) should be accounted, then the presence of excitons manifest in correlation function simply as an extra factor $ I_{\tilde{\theta}_{\sigma-}}(x)=\exp\left[(\imath\frac{K_{\sigma-}\gamma}{v_{\sigma-}})^2 \int^x d\Xi \int^x d\xi \langle a^{\dag}(\Xi-\xi/2)a(\Xi+\xi/2) \rangle \right] $ where a Deybe-Waller relation for a correlation of an exponential is used (and we moved to relative $\xi$ and global $\Xi$ coordinates of exciton)\footnote{The $\langle a^{\dag}(0)a^{\dag}(0) \rangle$ is projected out to high-energy sector $\omega\sim U$, the $\langle a^{\dag}(\Xi-\xi/2)a^{\dag}(\Xi+\xi/2) \rangle = 0$ in the absence of superfluid order parameter and the local bosonic density $\langle a^{\dag}(\xi/2)a(\xi/2) \rangle$ is assumed constant such that respective terms cancel}. Then to compute the
$I_{\tilde{\theta}_{\sigma-}}(x)$ we need to know a bosonic propagator in real
space or to be more precise a good approximation for an effective
propagator at low energies $\langle a^{\dag}(\xi) a(0)
\rangle_{eff}$. According to the previous section we take hard core bosons (that can be mapped on
spin-less fermions) and, following Eq.\ref{eq:eff-prop-ex} and $H_{\tilde{ex}}$, assume their dispersion is renormalized by
$V_{f-ex}$, such that $\omega_k\sim k$. Upon Fourier transformation of
a zero frequency limit $D_{ex}(k)\sim 1/k$ to real space we obtain (for a retarded Green's function) a Heaviside function
$D_{ex}^R(\xi)\sim \Theta(\xi-0)$. Then the causal function $\langle a^{\dag}(\xi) a(0)
\rangle_{eff}=\Theta(\xi-0)+\Theta(0-\xi)$. Substituting this above
we arrive at the following expression:
\begin{equation}\label{eq:I-theta-o}
I_{\tilde{\theta}_{\sigma-}}(x) = \exp\left[-\frac{K_{\sigma-}^{-2}\gamma^2}{v_{\sigma-}^2}\int_0^x d\Xi [(\Xi-x)+(\Xi+x)] \right]
\end{equation}
Upon performing the integration we arrive at:
\begin{equation}\label{eq:I-theta}
I_{\tilde{\theta}_{\sigma-}}(x) = \exp{\left[-\frac{K_{\sigma-}^{-2}\gamma^2}{v_{\sigma-}^2} x^2\right]}
\end{equation}
this factor $I_{\tilde{\theta}_{\sigma-}}(x)$ multiplies any correlation
function $\langle\exp\tilde{\theta}_{\sigma-}(x,t)\exp\tilde{\theta}_{\sigma-}(0,0)\rangle$. A very similar reasoning can be performed for a correlation
function of $\tilde{\phi}_{\sigma+}$ field, that leads to an analogous $I_{\phi_{\sigma+}}(x)$ factor. Overall the two Gaussians can be
combined and we arrive at a result that the fermionic spectral function, $A(x,t)=Im \langle\psi^{\dag}(x,t)\psi(0,0)\rangle$, has a form of
the TLL spectral function broadened by a Gaussian function:
\begin{equation}\label{eq:Ax-res}
  A(x,t)=\exp{\left[-\gamma^2\left(\frac{K_{\sigma-}^{-2}}{v_{\sigma-}^2}+\frac{K_{\sigma+}^2}{v_{\sigma+}^2}\right) x^2\right]}A_{TLL}(x,t)
\end{equation}
Please note that, exactly like in a case of Lang-Frisov
transformation for free fermions, the effect of excitons in TLL
enters through an exponential factor. From
Eq.\ref{eq:I-theta} we immediately deduce that in real space, for
a strictly local probe like STM, the effects of excitons are
invisible. Actually this is in agreement with the experiment, where a perfect fit to TLL was found in the STM\cite{Hager-bron-STM} and angle integrated PES\cite{Dudy-ARPES} measurements. To study a reciprocal space effects we need to make a
Fourier transform. We know the Fourier transforms for both
$A_{TLL}$ and a Gaussian, so in momentum space the total spectral function $A_{T}(q,\omega)$ is
simply $A_{TLL}(q,\omega)$ convoluted with a Gaussian:
\begin{multline}\label{eq:final-spectr-fun}
A_T(q,\omega)=\\
\beta^{-1}A_{TLL}(q,\omega)\otimes\exp\left[-\gamma^{-2}\left(\frac{K_{\sigma-}^{-2}}{v_{\sigma-}^2}+\frac{K_{\sigma+}^2}{v_{\sigma+}^2}\right)^{-1}q^2\right]
\end{multline}
Characteristic energy scale of a Gaussian is of the same order as
exciton-spinon coupling $\gamma$. A spectral function in the form $A_T(q,\omega)$ has been recently proposed\citep{Dudy-ARPES} to provide a very good fit to ARPES data of LMO in the low temperature phase, below 150K. Crucially, the maximal momentum broadening (indeed of order $\gamma/V_F$) was observed along the P-K line of BZ. Moreover, an additional bosonic bath introduces
an extra $\beta^{-1}$ factor. This is in agreement with a mysterious
experimental finding\cite{JWAllen-Tscal} where a perfect thermal scaling relation of TLL was found
with the only discrepancy that the thermal exponent $\eta$ was
shifted by $T^1$ (instead of $\eta=\alpha-1$, $\eta=\alpha$ was
detected). In the experiment\cite{JWAllen-Tscal} this new scaling seems to hold for spinons that are
coupled with the bath but not for holon part which seems to obey the "correct" TLL scaling without the
missing $T^{-1}$ power. Obviously more energetic many-body excitations, that is
those with $Max[\omega, T] >\gamma$, are to fast to be captured by
excitons, they move without the polaronic cloud. Hence the effects
of the broadening shall be visible only for sufficiently small
temperatures/frequencies. Another characteristic feature of our broadening is that its width does not depend on temperature.

An issue is whether the phase transition at 30K can be detected by
a measurement like ARPES. One could expect that the propagator of
localized excitons is quite different, which substantially
modifies Eq.\ref{eq:final-spectr-fun}. However one should notice that deep inside
the Wigner crystal phase the role of particles that carry on the
momentum (excitons) can be taken over by collective excitations –-
“phonons” of the Wigner crystal. The broadening Gaussian
changes its parameters, but a problem of how precisely it affects
an experimental signal is unclear and has to be be left for future studies.

\subsubsection{Influence on the RG equations}\label{ssec:RG-influence}

The appearance of the $I_{\tilde{\theta}_{\sigma-}}(x)$ and
$I_{\tilde{\phi}_{\sigma+}}(x)$ has very important consequences
for RG treatment of TLL in LMO. The fact that the correlation
function of one of the modes decays exponentially, implies that
all correlation functions that contains this mode are suppressed. So, for $\omega<v_{\sigma-}k_F$ where our derivation of Eq.\ref{eq:I-theta} holds, the $I_{\sigma\pm}$ factors will appear every time when the on-shell correlations are computed.
This affects beta functions for the non-linear interaction terms Eq.\ref{eq:umklapp-comb}-\ref{eq:2kF} that are
computed at every step of RG. The lower the on-shell energy scale,
the longer is the characteristic distance $x^{>}$ and the suppression is
stronger $I_{\tilde{\theta}_{\sigma-}}(x^{>}\rightarrow\infty)\rightarrow 0$. Due to this extra factor the renormalization group flow is slowed down already in
the \emph{intermediate} energy range. Each term that contains
either $\cos n\phi_{\sigma+}$ or $\cos n \theta_{\sigma-}$ (where
n is a real number) has the beta function that asymptotically goes to
zero (as if the cosine has effectively become a marginal perturbation). In particular, upon inspecting terms listed in p.5 in Sec.\ref{ssec:TLL-def} where these cosines are unavoidable, one realizes that this
mechanism prevents all SU(2) invariant Peierls distortions $V_P$ and $t_{\perp}$
hoppings (both terms are notoriously responsible for suppression
of 1D physics), from becoming violently relevant and destroying
TLL in LMO. This may provide an explanation for a particular
rigidity\cite{Dudy-ARPES} of 1D state in LMO.

It should be emphasized that the localization concerns the spin
sector of the theory, thus e.g. superconducting phases are not
prohibited. One should also note that the construction of
quarter-filling umklapp as well as $4k_F$ terms required
contractions of spin fields in OPE. These are local contractions,
with $x\rightarrow x'$, where
$I_{\tilde{\theta}_{\sigma-}}(x-x')\rightarrow 1$ and
$I_{\tilde{\phi}_{\sigma+}}(x-x')\rightarrow 1$, so these
constructions are not affected, $H_{um}$ and $H_{4k_F}^{\pi}$
remain well defined.

\subsection{Resistivity}

In order to explore the observable signatures of the phase
transition we need to investigate another probe: electrical
resistivity. Here the advantage is that excitons plays a very
different role below and above the transition.

Above the transition we discuss the problem of 1D fermions close
to quarter filling where the resistivity is caused by the $g_3$
umklapp terms. It has been postulated\cite{Piotr-bronze} that in
LMO the umklapp terms are marginal, hence on a verge of opening
the Mott gap. Conductivity can be expressed as:
$\sigma(\omega)=\imath v_{\rho+}K_{\rho+}/[\pi(\omega+M(\omega))]$
where $M(\omega)$ is a meromorphic memory function proportional to
a commutator $M(\omega)\sim[\Pi_{\rho+},H]$. The excitonic terms in Eq.\ref{eq:ham-disord-2}
do not contain coupling with $\rho+$ mode, hence they do not
influence the resistivity \emph{explicitly}. There are two cosine
terms in $H_{um}$ and $H_{4kF}$ that do contain cosines of
$\phi_{\rho+}$. We follow standard
procedure\cite{giamarchi_book_1d} to obtain the following
temperature dependence of DC resistivity (for quarter-filled chains):
\begin{multline}\label{eq:rho-T}
  \rho(T)= \left(\frac{g_3^2}{\pi b^{-1} \sqrt{v_{\rho+}}}\right) \left(\frac{2\pi b^{-1} T}{v_{\rho+}}\right)^{(8 (K_{\rho+}+K_{\rho-})-3)}\\
  (B(2(K_{\rho+}+K_{\rho-}),1-4(K_{\rho+}+K_{\rho-}))\cos(2\pi (K_{\rho+}+K_{\rho-})))^2\\
  +\left(\frac{g_{4kF}^{2}}{\pi b^{-1} \sqrt{v_{\rho+}}}\right) \left(\frac{2\pi b^{-1} T}{v_{\rho+}}\right)^{(16 K_{\rho+}-3)}\\
  (B(4K_{\rho+},1-8K_{\rho+})\cos(4\pi K_{\rho+}))^2
\end{multline}
a formula which can be checked against experiments. For $K_{\rho+}\approx 0.25$ and $K_{\rho+}\approx 0.35$, values predicted in Ref.\onlinecite{Piotr-bronze}, we expect that the dominant contribution will be $\rho(T)\sim T^{1.7}$ and the sub-dominant will be $\rho(T)\sim T^{1}$. Based on arguments from Sec.\ref{ssec:RG-influence}
we have excluded the terms in $H_{um}$ proportional to $\xi$, as they will be
affected by excitons, but naively there should be no further effect of the
excitonic clouds.

What is neglected in this reasoning are initial (UV-RG) amplitudes
of the non-linear interaction terms. The strength of a bare coupling
can be computed by a standard prescription of going from first to
second quantization -- to be precise we need an overlap between an
excitonic wave-function $\psi_a(r)$ and wave-functions of
interacting electrons. If we use Lang-Frisov transformation,
Eq.\ref{eq:Lang-Frisov}, then this physical description takes the
following mathematical form:
\begin{multline}\label{eq:g3-ampl-1}
\tilde{g_3}= \int dx_2 \int dx_1 \int dx \bar{U} \Box((x-x_1)/b)\Box((x_1-x_2)/b)\\
\langle\exp[\gamma/\omega_{k_F} \psi_a^{*}(x)\psi_a(x)
]\rangle\langle\rho_{xy}(x_1)\rangle\langle\rho_{xy}(x_2)\rangle
\end{multline}



where $\bar{U}=U+V(2_{k_F})$ is an effective strength of local umklapp potential, $\Box(x)$ is a rectangle function with a width two (this ensures local character of exciton-fermion state and fermion-fermion interaction), $\langle\rho_{xy}(x_1)\rangle$ is an expectation value of the $d_{xy}$ fermion density and $\psi_a^{*}(x)$ is a wavefunction od the $a$-particles, excitons (in many body language the relation between $a^{\dag}$ and $\psi_a^{*}(x)$ is the same like between $c^{\dag}(x)$ and $\psi(x)$). 
By Taylor expanding the exponential we arrive at the following
correction to $g_3$:
\begin{equation}\label{eq:g3-ampl-1}
\Delta\tilde{g_3}\approx \frac{\bar{U}n_{ex}}{2\pi}\left(\frac{\gamma}{\Gamma}\right)^2 \int_{-b}^{b} d x \langle
\psi_a^{*}(x)\psi_a(x)\rangle ^2\langle\rho_{xy}(x)\rangle\langle\rho_{xy}(x)\rangle
\end{equation}
where $n_{ex}$ is a number of excitons and a factor $1/2\pi$ comes from integrating two presumed Gaussians (over $x_1,x_2$). A similar correction can be derived for the $g_{4kF}$ as well. 
To evaluate an integral Eq.\ref{eq:g3-ampl-1} one needs a precise
form of a wavefunctions which makes it a hard task. However,
simply by noticing that both $\psi_a(x)$ (see App.\ref{ssec:resonant-state}) and $d_{xy}$ have the
same node-less character $l_z=0$ along a common b-axis, one can
deduce that there is no cancellation by symmetry and the integral
is finite.


We arrive at a new class of problem where for a randomly chosen
\emph{fraction} of fermions amplitudes of umklapp scattering are modified. This is,
in essence, a disorder put on the top of Mott physics with $g_3(x)$ changing randomly in space/time. From
numerical studies of a standard Mott-Anderson problem it is known that the two localization mechanisms compete and the critical $U$ shifts to
larger values in the presence of disorder. Our, slightly modified problem, is an extremely interesting research direction, which needs to be postponed to further studies.

Below $T_o$ excitons form a crystal, so the excitonic clouds
disappear. There is no randomness or suppressed RG beta functions,
moreover interactions changes character due to presence of a
crystal of dipoles: from Coulomb with $V_{Coul}\sim 1/q$ to (at
most) dipolar with $V_{di}(q)\sim F[1/r^3]\sim cste(|q|)$, that is
a constant independent of $|q|$. While all these factors can push the $d_{xy}$ system into the Mott insulator phase, they are of only minor importance, since at temperatures of order 40-50K a presence of a finite doping $\delta$ should dominate the flow and prevent any upturn of resistivity. Once the
system's energy is lowered enough, such that it realizes the
finite doping $\delta$, the umklapp processes should be suppressed.   


This contradiction with experimental findings can be re-solved thanks to ordering of excitons. Due to
the emergence of the Wigner crystal with periodicity $\pi
q_W^{-1}$ along the b-axis\footnote{$q_W$ is not necessarily directly
related to concentration of excitons: a dilute excitons may form
2D Wigner crystal which is tilted with respect to a structural
crystal lattice, then (by an overlap of the two lattices) de Moire
pattern will form with a non-obvious periodicity along the
selected b-axis} excitons acquire a finite expectation value for a
finite value of momentum $q_W$, that is $\langle a_k^{\dag}
a_{k+q_W}\rangle = \Psi_W \neq 0$. Thus there is a new term when
any correlation function over the excitonic cloud is computed $\langle \psi_a^{*}(x)\psi_a(x)\rangle
\sim \langle a_k^{\dag} a_{k}\rangle + \exp(\imath q_W x) \langle
a_k^{\dag} a_{k+q_W}\rangle$.
Following Eq.\ref{eq:g3-ampl-1} this leads to an additional
correction to the umklapp amplitude $\Delta g_3 \sim |\Psi_W|
\cos(q_W x)$, where $|\Psi_W|$ is an order parameter of the
excitonic crystal. As a result we expect a following term $\sim
|\Psi_W|\cos(\phi_{\rho+}+x(\delta-q_W))$ to appear in the
hamiltonian. Clearly, the periodicity of Wigner crystal may
compensate the effect of a finite doping and cure the problem of
incommensurability. Importantly the effect is proportional to the
order parameter $|\Psi_W|$, which implies that it is enhanced when
the temperature is reduced.

To be precise, in AF-Ising model on the trigonal lattice,
immediately below $T_o$ one expects\cite{Moessner-QIsing} an
intermediate BKT phase with a quasi-long range order. A simplistic
picture of this order, a toy-model to capture a mechanism of resistivity just below $T_o$, is in the first step formation of pairs of excitons
separated by a distance $\pi q_W^{-1}$.  A physical interpretation
of resistivity would be that each of these pairs of excitons (like
impurities) has a holon locked in between them on a "quantum dot"
with a small mass term $\sim T_o$ corresponding to a quantum
capacitance. Such problem, taken together with a cosine potential
(from the rest of $H_{um}$ in our case), has been discussed
extensively in Ref.\onlinecite{giamarchi_book_1d} (Chap.10.2.3).
One defines "local" fields
$\phi_{\rho+}^{\pm}=1/2[\phi_{\rho+}(-\pi q_W^{-1}/2)\pm
\phi_{\rho+}(\pi q_W^{-1}/2)]$ for which the action is $S_0=\sum
|\omega_n|/K_{\rho+}
\phi_{\rho+}^{+*}(\omega_n)\phi_{\rho+}^{+}(\omega_n)$ plus the
capacitance term from $\phi_{\rho+}^{-}$. With this the problem
can be mapped onto tunneling through strong impurity. Based on
this mapping, we are able to predict the following temperature
dependence of resistivity:
\begin{equation}\label{eq:rho-T-low}
  \rho(T<T_o)\sim T^{2-2/(2 K_{\rho+})}
\end{equation}
hence in this toy-model of cooperative localization the resistivity shall increase as $\rho(T<T_o)\sim T^{-1.7}$ (again for $K_{\rho+}\approx 0.27$). This falls close to an experimentally measured value\cite{Dudy-ARPES,Santos-transport-weird}.

\section{Discussion}\label{sec:discuss}

Naturally, we would like to explore if there are other experimental probes that can provide us a clear hallmark of the novel physics.

Firstly, we discuss an influence of an external magnetic field on a resistivity in the low temperature phase. In other words we discuss the influence of the magnetic field on the order parameter $|\Psi_W|$. The most astounding property of dark excitons, their ultra-long recombination time, is linked to angular momentum conservation. This can be spoiled by applying an external magnetic field $\vec{B}$ perpendicular to the quantization axis. For both terms $H_{f-ex}^{(1,2)}$ one could introduce quantization axis parallel to a-axis (this makes $\langle J_z \rangle$ maximal), so spin-states mixing will be present for $\vec{B}$ along the b or c axis. It is known\cite{Rodina-dark-recomb} that the "dark excitons", thus our hard-core bosons, acquire a finite life-time $\sim B^2$. Then they leak out of the system and one should observe melting of the Wigner crystal. A suppression of resistivity will follow. This goes in-line with experimental findings from Ref.\onlinecite{Hussey-rhoB}. The negative magneto-resistance effect when $B||b$ is monotonous and saturates close to 18T. This energy scale corresponds to 20K that is fully melted crystal, $|\Psi_W|\rightarrow 0$, so no contribution to resistivity. When $B||c$ the situation is more complex: it seems that there are two competing mechanisms and Wigner crystal melting dominates only at high magnetic fields. At low magnetic fields fermions' movement along the b-axis couples with the magnetic field which gives rise to Landau diamagnetism and a standard positive magneto-resistance.

Excitons are usually associated with optical probes, but "dark excitons" do not couple with light, so they are invisible by standard spectroscopic methods. This can be avoided by applying an external magnetic field which, as already mentioned, mix various orbital momenta and hence mix bright and dark excitonic states. (In our case the energy gap caused by excitonic crystal formation $\Delta E \approx 20K$ prohibits this mixing, thus $\Delta E$ competes with the magnetic field). We deduce that excitonic states should provide visible effects in magneto-optical spectroscopy for sufficiently large magnetic field. Then they can absorb light in a process when electron is transferred to other d-orbitals (previously prohibited due to $\tilde{L}$ conservation). Indeed, this allows to interpret a magneto-chromatic effect detected in Ref.\onlinecite{Choi-colors} , with characteristic energy scale $\approx 20K$, where authors concluded that magnetic field seems to cause “reorganization of d-orbitals of an unknown origin”. Remarkably, the effect is present only for $B||b$ in a very close resemblance to magneto-resistivity discussed above.


Finally, let us describe phenomena related to the heat transfer. Obviously excitons, as a new dynamic degree of freedom in the system, do carry some heat capacity. For a linear dispersion, this heat capacity scales like $\sim T^3$. An issue is an amplitude of this signal: number of excitons is small and they are the least mobile component of the system. Moreover in the high temperature regime they are bounded to $d_{xy}$ fermions, so one needs to disentangle the two signals. The heat carried by excitons may be hard to distinguish e.g. from acoustic phonons. However, a detectable effects are expected in magneto-thermoelectric measurements, a transverse signal induced by a magnetic field, preferably with a magnetic field along the c-axix which does not lead to an immediate extinction of excitons. We should then measure a tunneling signal in between the slabs (along the a-axis) induced by an electric current along the b-axis. The Lorentz force acts on each electron and also on an accompanying exciton cloud. For LMO, which is in the regime $K_{\rho\pm}<1/3$, the particle-hole tunneling is more relevant than single particle tunneling between TLLs. Then the excitonic contribution can be substantial and distinguishable as it carries the heat, but does not carry the charge. This may offer an explanation of gross violation of Wiedmann-Frantz ratio detected in Ref.\onlinecite{Hussey-thermo}.

\section{Conclusions}

In conclusion, the main result of this work is to incorporate, in a concise manner, a multi-orbital physics on the top of a well established, many-body Tomonaga-Luttinger liquid construction. Our idea can be understood as an effort to capture an entanglement (also non-local) between various d-orbitals within the $t_{2g}$ manifold. A model that we obtained, based on excitonic physics, is able to explain several experimental observations in
purple bronze, that were so far impossible to reconcile. This includes: Gaussian deviations from TLL spectral function detected by ARPES\cite{Dudy-ARPES}, missing power of temperature in the scaling relation for spinons\cite{JWAllen-Tscal}, mysterious upturn of resistivity that is extremely fragile upon applying magnetic field\cite{Hussey-rhoB} and magneto-chromatic effect\cite{Choi-colors}. We also made predictions for magneto-thermoelectric coefficients. An intriguing path, that has never been even anticipated, is now laid down for further studies of this exceptional
material. Without claiming to make a final step in understanding
all properties of LMO, certainly a significant amount work is
going to be necessary, we believe that future experimental findings should be
at least cross-examined in a view of our proposal.

The ideas put forward in this paper have also some further reaching consequences. For the material under
consideration, they pave the way to explore the origin of
superconducting phase that is present at even lower temperatures. Current work
substantiates the link between 1D physics and SC order. For the
general problem of dimensional cross-over, present in every
quasi-1D material (or structure), it emphasizes so far neglected
role of multi-orbital physics, when the system may cross-over in
an orbital-selective manner. We also gain knowledge about Mott transitions as LMO provides an illustrative example where auxiliary degrees of freedom may help the system to achieve
commensurability and open up a gap.

[Note added: During final preparation of this manuscript, the author has become aware of a very recent experimental work, Ref.\onlinecite{Clark-magnetic-moment}. The picture put forward in the conclusion of this paper, about the presence of fluctuating magnetic moments on atomic scale that change their character at $T_o$, goes exactly along the lines of theoretical ideas put forward in current paper.]

\section*{Acknowledgments}

It is a pleasure to thank several people without whom this work would be impossible. Firstly, I wish to thank Thierry Giamarchi for many inspiring discussions about forward scattering in TLL. I would like to thank Jim Allen for sharing his profound knowledge about LMO. Finally, I want to thank Lenart Dudy for patient explanation of all peculiarities of ARPES experiments.

\appendix

\section{Spin-orbital content of an exciton}\label{ssec:resonant-state}

Since the spin-orbit coupling is substantial ($\Delta_{LS}=100meV$ on a molybdenum atom\cite{Iverson-Mo-LS}), in order to understand excitonic wavefunctions we need to
consider eigenstates of total angular momentum $J$. The
case $J=0$ is the simplest (nodeless), while $J=2$ is the most complicated.
In an isotropic system, without a crystal lattice, a state with an
angular momentum $J=2$ would be five-fold degenerate. One should
realize that our system is not only highly anisotropic, with a
preference for wavefunctions to be spread within the b-c plane,
but also unidirectional in the sense that each electron/hole have
one direction along which they strongly prefer to
move\cite{Popovic-bron-DFT, Canadell-DFT-old}. [Please note that
here we discuss the wavefunction that corresponds to relative
motion of electron and hole that are forming an exciton]. This
clearly manifests in ARPES where three sets of unwrapped bands
were observed\cite{JWAllen-old}\footnote{Jim Allen's private
communication}, each for a different $t_{2g}$ orbital. In such a
case the wavefunctions should have an elongated shape, very much
like $J_z=0$ spherical harmonics. To be precise there are two
possibilities of preferred orientations of these rods one for
$d_{xz}$ and another for $d_{yz}$. Mathematically a difference
between them can be ascribed to different phases of an ad-mixed
$|L_z|=1$ component. A naive guess would be that either one of
these two (or some linear combination that favours an intermediate
direction) would eventually become a ground state. However, in the
hamiltonian there are no symmetry breaking terms, instead there
may be substantial disorder (octahedron tilting and dynamic
coupling to $d_{xy}$ orbitals) that restores the symmetry. Then we
exclude spontaneous symmetry breaking towards these unidirectional
"stripe-states". On the other hand, when the exciton's energy is
close to $E_F$ then $L_z=0$ (the $d_{xy}$ states) are available at
low energy cost as well as $L_z=\pm 1$ so one can act with
$\tilde{L}^{\pm}$ operators with no energy cost. Then an isotropic
ground state is expected to be resonance of many Fock states
(tensor products) spanned over eight available Hilbert states (it
can be thought as a purely \emph{local} analogue of the famous RVB
state). [Two states are from rods orientation, two from
electron/hole (Nambu) degree of freedom and two from spin.]  Most
likely, in a low symmetry environment of LMO, there is only one non-degenerated excitonic state $a(x)$, but it
is unusual since the $J_z$ fluctuations are incorporated in its
construction. The Holstein-Primakoff mapping (used also
Sec.\ref{ssec:excit-propert}) indicates that it is possible,
creation of boson is equivalent to $J_{+}$ operator. Physically,
creating an exciton implies a creation of the fluctuating
orbital-state.

\bibliography{bronze-LL}

\end{document}